# One-dimensional quasi bound states in the continuum with large operation bandwidth in the ω~k space for nonlinear optical applications

**Kaili Sun**[1,#], **Hui Jiang**[1,#], **Dmitry A. Bykov**[2], **Vien Van**[3], **Uriel Levy**[4], **Yangjian Cai**[1] and **Zhanghua Han**[1,*]

*1. Shandong Provincial Key Laboratory of Optics and Photonic Devices, Center of Light Manipulation and Applications, School of Physics and Electronics, Shandong Normal University, Jinan 250358, China*

*2. Image Processing Systems Institute — Branch of the Federal Scientific Research Centre "Crystallography and Photonics" of Russian Academy of Sciences, 151 Molodogvardeyskaya st., Samara 443001, Russia*

*3. Department of Electrical and Computer Engineering, University of Alberta, Edmonton T6G 2V4, Canada*

*4. Department of Applied Physics, and the center for Nanoscience and Nanotechnology, The Hebrew University of Jerusalem, Jerusalem, Israel*

*\*zhan@sdnu.edu.cn*

*# These authors contribute equally to this work.*

The phenomenon of bound state in the continuum (BIC) with infinite quality factor and lifetime has emerged in recent years in photonics as a new tool of manipulating light-matter interactions. However, most of the investigated structures only support BIC resonances at very few discrete points in the ω~k space. Even when the BIC is switched to a quasi-BIC(QBIC) resonance through perturbation, its frequency will still be located within a narrow spectral band close to that of the original BIC, restricting their applications in many fields where random or multiple input frequencies beyond the narrow band are required. In this work, we demonstrate that a new set of QBIC resonances can be supported by making use of a special binary grating consisting of two alternatingly aligned ridge arrays with the same period and zero-approaching ridge width difference on a slab waveguide. These QBIC resonances are distributed continuously over a broad band along a line in the ω~k space and can thus be considered as one-dimensional QBICs. With the Q factors generally affected by the ridge difference, it is now possible to choose arbitrarily any frequencies on the dispersion line to achieve significantly enhanced light-matter interactions, facilitating many applications where multiple input wavelengths are required, e.g. sum or difference frequency generations in nonlinear optics.

1. **Introduction**

The concept of bound states in the continuum (BIC) was first proposed in quantum mechanics by von Neumann and Wigner in 1929[1], predicting the existence of localized eigenstates of the single-particle Schrödinger equation embedded in the continuum of eigenvalue state solutions. This counterintuitive observation is of fundamental importance in quantum mechanics. Over the years, the phenomenon of BIC has also been popularized and extensively studied in various fields, like acoustics[2,3], electronics[4–6] and microwaves[7,8]. In 2008, the concept of BIC was further extended to optical systems for the first time by D. C. Marinica and A. G. Borisov[9], and has ever since become a new approach of enhancing light-matter interactions. The main idea of BIC is the elimination of the coupling between the resonant modes and all radiation channels in the surrounding space. There are two main approaches of achieving BICs. The first is the so-called symmetry-protection, which generates BIC at the $\Gamma$ point in the reciprocal space and it is based on the symmetry incompatibility between the bound state and the continuum. A bound state of one symmetry class can be embedded in a continuum of another orthogonal symmetry class, and their coupling is forbidden when there is a zero overlap integral between the modes of different symmetry properties. The other type is due to the accidental disappearance of the coupling coefficient with the radiated waves by a successive tuning of one or more system parameters resulting in the formation of the so-called accidental BIC, which is typically observed away from the $\Gamma$-point and is therefore also known as off-$\Gamma$ BIC. The accidental BICs can be explained by the destructive interference of two or more leaking waves, where the radiation from the leaking waves is tuned to cancel each other out completely, a mechanism also known as the Friedrich-Wintgen scenario[10]. Two resonances are usually involved in the accidental BIC and due to the strong coupling between the two resonances, an avoided crossing behavior is usually accompanied[11].

In optics and photonics, researchers have realized both symmetry-protected BICs and accidental BICs, as well as QBIC resonances with finite Q factors and easier excitation requirement by introducing some geometrical perturbations, in a variety of structures including gratings[11–13], waveguide arrays[14,15], photonic crystals with near-zero refractive index[16,17], integrated photonic circuits[18,19] and metasurfaces[20,21]. A range of amazing properties and applications have been achieved to date with photonic QBICs in applications such as lasing[22], sensing[23] and Raman spectroscopy[24], and the application of QBIC in nonlinear optics[25], twisted light[26] and light-matter interaction[27] is being actively investigated. Unfortunately, most of the reported BICs only occur at very few discrete points in ω~k space for a given structure geometry. This is true for both the symmetry-protected and accidental types of BIC. When the ideal BIC is switched to a QBIC resonance, its frequency remains within a narrow spectral band tightly close to that of the original BIC, even at a highly different k wavevector away from the BIC point in the ω-k space. As a result, the enhanced light-matter interactions can only be achieved within the narrow band for a specific geometry, restricting the applications of BIC in many circumstances where it is required to tune the working frequency or to have multiple inputs beyond the band. Due to this reason, most of the BIC applications in nonlinear optics reported to date are focused on higher harmonic generations [28,29], where only a single light beam at the fundamental frequency is involved. The ultra-high local electric field enhancement associated with the QBIC resonances significantly improves the nonlinear conversion efficiency [25]. However, there are many circumstances in the nonlinear optics where the generation of the target signal frequency requires two or more different input frequencies, e.g. in sum frequency generation (SFG) or difference frequency generation (DFG). To obtain the utmost enhancement based on the BIC effect, one needs to have all the input frequencies at the QBIC resonances. Unfortunately, one can see that significant restrictions emerge with the conventional QBIC because in those applications, the input frequencies may not be within the narrow operation band of QBIC resonance supported by a structure. The situation becomes even worse if one needs to realize a spectral tunability of the target signal frequency.

In this work, we demonstrate that a binary grating composed of two ridge arrays with the same period and different ridge width on a slab waveguide structure supports a new set of QBIC resonances and can address the above challenges. We note that a similar structure with regular uniform ridge grating supports the ideal BIC at normal incidence and the QBIC at inclined incidence, both at fixed and limited number of frequencies[12]. In contrast, these QBIC resonances supported by the binary grating are distributed continuously along a line over a large spectral range in the ω~k space and thus can be considered as one-dimensional (1D) QBIC. Using the 1D QBIC supported by a structure with fixed geometry, it is now possible to choose arbitrarily any frequency within a broad range to achieve enhanced light-matter interactions. This important feature greatly promotes many applications requiring multiple input wavelengths, such as SFG or DFG, and is expected to significantly push forward the use of BIC in applications such as nonlinear optics.

## 2. Structure and results

### A. Dispersion and Q factors

Figure 1 shows a schematic sketch of the investigated structure that supports the 1D QBICs. The red dashed box in the inset shows a magnified side view of the unit cell. To demonstrate the working principle, Si (dark grey area, refractive index 3.45) is first assumed in this section as the constituent material for both the slab waveguide and ridges on the substrate of $SiO_2$ (blue area, refractive index 1.45). To achieve the QBIC effect in the telecom band, we deliberately adjusted the period P to 540 nm, and chose t = 220 nm as in the standard silicon on insulator (SOI) wafer specification. For the topmost grating layer, we take the width of one ridge as w=50nm and define another as w+δ by introducing the deviation variable δ, both with a height h of 60nm. For the sake of simplicity and the ease of calculation, it is assumed that all materials are dispersionless. The red arrows in Fig. 1 represent the incident, reflected and transmitted light beams. Without losing generality, we consider only TE-polarization with the electric field parallel to the y-direction and the incident beam is within the *xz*-plane onto the structure at an angle of θ with respect to z axis. Similar results and conclusions can be obtained as well for the TM-polarization. A finite-element method based commercial software of Comsol Multiphysics together with Floquet periodic boundary conditions in the x-direction and perfectly matched layers (PML) in the z-direction is used for all the calculations.

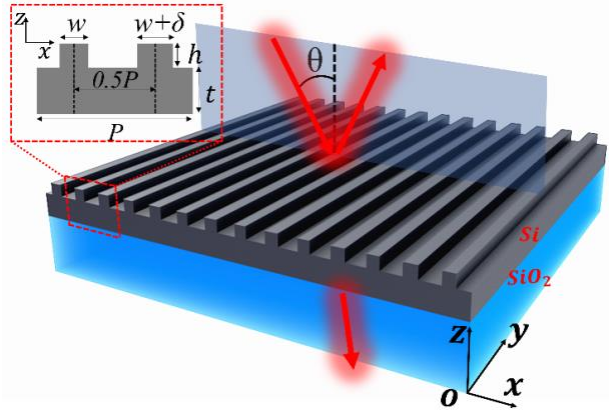

Fig.1. Schematic diagram of the structure supporting the 1D QBICs. The inset presents a magnified view of the grating unit cell, which is assumed to extend infinitely along the y-direction. The red beams indicate the incident, reflected and transmitted light.

When the ridge width difference δ is non-zero, the binary grating has a period of P to accommodate two ridges within one unit cell and the well-known phenomenon of guided-mode resonance (GMR)[30] is achieved, exhibiting a sharp dip in the transmission spectrum when the following phase-matching condition is satisfied:

$$k_0 \sin\theta + m\frac{2\pi}{P} = k_0 n_{eff} \qquad (1)$$

where θ is the incidence angle, $k_0$ is the wavenumber in vacuum, $n_{eff}$ is the effective index of guided mode within the slab waveguide and *m* is the order of diffraction by the grating. For simplicity, we only consider the ±1 diffraction orders in this paper. To demonstrate that the binary grating structure with a non-zero δ supports the QBIC resonances, we show first in Fig. 2(a) the calculated dispersion properties of the resonances in the ω~k space with a special value of δ as 10nm. The dashed line represents the dispersion of light in air, above which it is the continuum region of radiation. One can also observe that there are two branches of resonances in Fig. 2(a), which arise from the excitation of guided modes counter-propagating towards two differenet directions in the waveguide. The relative position of those resonances above the light line confirm the leaky characteristics of these resonances. We are more interested in the high-frequency branch of the two bands, which has a resonance at the Γ point (as illustrated by the dashed circle in Fig. 2(a)) with a Q factor of infinity. Keeping δ as 10nm, we further present in Fig. 2(b) the dependence of the calculated Q factor on the incident angle for resonances on this branch. It is found that the Q factor increases significantly with the decrease of the incident angle and approaches infinity at normal incidence, suggesting the state of an ideal BIC. Considering the symmetry of the binary grating cross the central plane of either ridge, we believe this ideal BIC resonance belongs to the symmetry-protected type. The finite yet ultrahigh level of Q factors at non-zero incident angles suggest that other resonances on the same branch can be considered as QBIC resonances. In contrast to conventional QBIC resonances which are due to some structural perturbations introduced into the geometry, these QBIC resonances result from the perturbation in the incident angle instead. We should note that the spectral positions of these QBIC resonances are still governed by equation (1). As a result, these QBIC resonances have a dispersion curve which covers a much larger bandwidth, giving rise to the terminology of 1D QBIC resonances. For the conventional QBIC resonances supported by resonating metasurface elements, although a continuous distribution of ω~k dependence can also be found [25,29], the dispersion curve is much flatter within a much narrower spectral band. We note that the other branch of low-frequency resonances shown by the red line in Fig. 2(a), although still representing leaky modes with ultrahigh Q factors, are not referred to as the QBIC resonances herein because they don't originate from the perturbations either in the incident angle or in the structural geometry. As a result, only the resonances on the high-frequency branch will be discussed hereafter.

The fact that the spectral positions of 1D QBIC and the corresponding incident angles are still determined by equation (1) suggests that these 1D QBICs can also be considered as GMRs. It is well-known that the regular GMR effect themselves can generate a sharp resonance effect with large Q-factor [30]. To have a straightforward comparison between the GMR and the 1D QBIC effects supported by a conventional ridge grating and the binary grating respectively, we present in Fig. 2(c) the transmission spectrum at the same incident angle 2° for both the GMR (red line) and the QBIC (black line) with the inset showing the structure schematics of the two cases. Here δ is still chosen as 10 nm for the binary grating. It is clearly shown that a much sharper transmission dip is presented for the QBIC resonance than the GMR. Detailed calculations show that the QBIC resonance has a Q-factor higher than $10^4$ and is two orders of magnitude higher than that of the GMR shown in the same figure.

The above results are obtained with a δ value of 10nm. Actually the general Q factor of the 1D QBIC resonances over a large bandwidth can be significantly affected by the level of asymmetry between the two ridges. The dependence of resonance Q factor as a function of δ at two random incident angles of 3° and 5° are calculated and presented in Fig. 2(d). It is quite evident that the general Q factor increases with the decrease of the ridge asymmetry. A polynomial fitting of the Q values as a function of δ are shown in the inset of Fig. 2(d) and an inversely quadratic dependence of Q versus δ is found, in a similar way as the QBIC resonances supported by asymmetric metasurfaces[31]. Actually, the same behavior are found to work for any incident angle. When δ is non-zero yet small, the 1D QBIC resonances with finite yet ultra-high Q factors can be excited. One can choose any wavelength within a broad spectral range to have the QBIC resonance, whose Q factor can be further controlled by choosing a proper deviation of δ from 0. These 1D QBIC resonances are of special importance for real applications, thanks to the benefit of relieved excitation requirement while the local enhancement of electromagnetic fields is weakly affected.

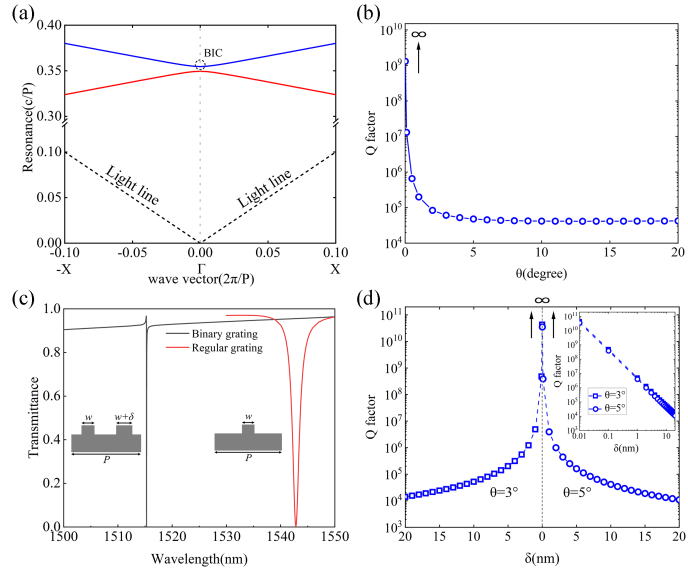

Fig. 2. (a) Band structure of the BIC/QBIC mode supported by the binary grating waveguide structure when δ is 10nm. The dotted line represents the light line in the free space. (b) Q factor as a function of incident angle for the QBIC resonances at δ=10 nm. (c) The transmission spectra of a regular grating structure GMR (red line) and binary gratings (black line) both at incident angle of 2°. (d) Dependence of resonance Q-factor on the ridge width difference δ, at two incident angles of 3°(diamond) and 5°(circle) respectively.

### B. Coupled-mode theory.

The above behavior in Fig. 2 can be actually theoretically explained within the framework of coupled-mode theory. As we described above, the Q factor of the QBIC depends on both the incidence angle θ and ridge width difference δ. Here we adopt the spatiotemporal formulation of this theory for gratings presented in paper [33], and write the homogeneous coupled-mode equations:

$$\begin{cases} \dfrac{\partial u}{\partial t} = -v_g \dfrac{\partial u}{\partial x} + c_1 u + c_2 v; \\ \dfrac{\partial v}{\partial t} = v_g \dfrac{\partial v}{\partial x} + c_1 v + c_2 u. \end{cases} \qquad (2)$$

Here $u$ and $v$ define the amplitudes of the two counter-propagating modes of the slab layer; $v_g$ is the group velocity of these modes, and $c_1$ and $c_2$ are the coupling coefficients. Making use of the energy conservation law[34] we can show that $i(c_1-c_2)$ is real for lossless structures.

Taking the Fourier transform of Eq. (2) we can arrive at the system of linear equations having non-trivial solutions when

$$v_g^2 k_x^2 = (\omega - \omega_{p1})(\omega - \omega_{p2}), \qquad (3)$$

where $k_x = k_0 \sin\theta$ and $\omega = 2\pi c/\lambda$ are the wave number and angular frequency of the incident light; $\omega_{p1} = i(c_1+c_2)$ is the complex frequency of the symmetric mode at the bottom of the red line in Fig. 3(d), and $\omega_{p2} = i(c_1-c_2)$ is the *real* frequency of the antisymmetric mode marked by (a). Equation (2) is the dispersion equation describing the hyperbola-like dispersion law seen in Fig. 3(d).

Note that all the parameters used in Eqs. (2) and (3) depend on the ridge width difference δ. For the considered structure it is the dependence of $\mathrm{Im}(\omega_{p1})$ on δ is the most important. As we demonstrated previously, the eigenmodes of the structure are not exited at δ = 0. Therefore, the frequency $\omega_{p1}$ of the symmetric mode is real at δ = 0 and $\mathrm{Im}(\omega_{p1}) \approx \delta \cdot \alpha$ where $\alpha$ is a real parameter. When δ is zero all coefficients in Eq. (3) become real and we obtain the dispersion law for the 1D QBIC: $v_g^2 k_x^2 = (\omega - \mathrm{Re}\,\omega_{p1})(\omega - \omega_{p2})$. When the ridge width difference is non-zero, we can solve Eq. (3) with respect to the complex frequency ω and obtain the Q factor $Q = \mathrm{Re}\,\omega / (-2\,\mathrm{Im}\,\omega)$ of the QBIC.

*C. Transmission spectra.*

We further investigate the properties of the 1D QBIC resonances, by studying the transmission spectra of the binary grating at different incidence angles. For the binary grating structure with δ=10nm, in Fig.3 (d) we present the calculated resonance wavelength versus incident angle. When the incident angle increases, the high-frequency branch undergoes a blue shift, while the other branch experiences a red shift to the opposite direction. These trends suggest that the low-frequency branch results from the grating excitation with $m = -1$ while the other branch is with $m = 1$ in equation (1). We choose three points in Fig. 3(d) marked a ~ c to demonstrate the formation of BIC/QBIC modes for different cases, where a corresponds to the resonances at normal incidence, while b and c correspond to two randomly selected incident angle of 3.5° (b) and 5° (c). We present both the transmission spectrum and the field distributions around these three points in Figs. 3(a) ~ (c) respectively.

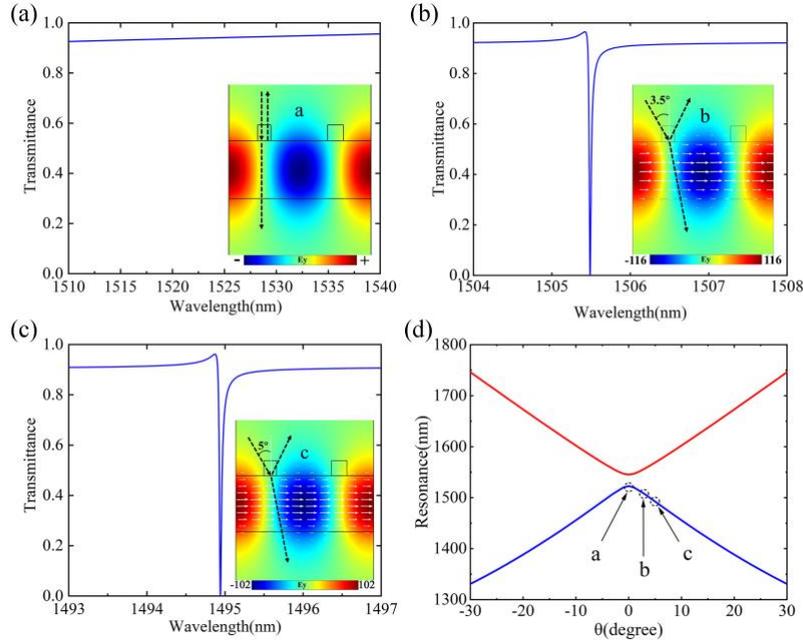

Fig.3. (a)~(c) present the local transmission spectrum close to the three positions marked in Fig.3 (d), with the inset showing the field distribution of the real part of $E_y$ (the inset of (a) is obtained from eigen-frequency analysis) and the white arrows representing the vectorial distributions of the Poynting vector. The black arrows in each figure indicate incidence, reflection and transmission respectively. (d) The relationship between the resonant wavelength of BIC/QBIC modes and the incident angle in the binary grating structure. The three circles of a ~ c represent the BIC/QBIC modes at different angles.

Point a has a completely different property as compared to other points. Eigen frequency analysis demonstrates that a resonance with an infinite Q factor can be found at this point. The transmission spectrum calculations also show that this resonance cannot be excited by a plane wave at normal incidence (see Fig. 3(a), where the mode distribution is obtained from eigen frequency analysis). These results suggest that point a corresponds to the occurrence of an ideal BIC resonance. Although there is a slight level of structural asymmetry between the ridges (δ=10nm), the whole structure is still symmetric if one uses a vertical plane cross the center of either ridge. So the resonance at point a is an ideal BIC of the symmetry-protected type, and thus cannot be excited by a plane wave at normal incidence. It is seen from the inset of Fig.3(a) that the electric field is mainly concentrated within the waveguide layer and located between the grating ridge intervals. So it is distributed with perfect anti-symmetry. Judging from the same E field distributions at higher incident angels (cf. the inset in Figs. 3(a), (b) and (c)), we believe any other points along the same band away from position a can be interpreted as QBIC resonances which result from a perturbation in the incident angle from the symmetry-protected BIC at position a. As a result, the QBIC resonance in this high-frequency branch typically has a Q factor which decreases with the increase in incident angle. For a larger incident angle, the wavelength of the QBIC has a blue-shift. For example, when the incident angle is 3.5°/5°, the QBIC wavelength decreases from the BIC wavelength of 1522.8nm to 1505.5nm/1494.9m, as marked by points (b) and (c). Since a larger incident angle means higher perturbation from normal incidence, the corresponding Q-factor drops slightly, consistent with the result in Fig. 2(b). As is shown in the inset of Figs. 3(b) and (c), the power flow propagates from the left to the right side of one unit cell, along the same direction as the incident beam, which corresponds to the grating diffraction order *m* to be 1.

### 3. SFG as an example of application

To demonstrate that it is possible to select randomly any incident wavelength within a specific range and generate the light in the desired spectral range with a larger freedom of choice by using this 1D QBIC resonances, we choose the process of SFG as a simple example. Here an x-cut (the optical axis is along y direction) $LiNbO_3$ film structure is employed to make use of its relatively high second-order nonlinear susceptibility along the TE polarization. The structure is schematically shown as the inset of Fig. 4(a) and its geometrical parameters are adjusted due to a smaller refractive index ($n_o$=2.22, $n_e$=2.14) of $LiNbO_3$ compared to Si. Fig. 4(a) presents the calculated transmission spectra for TE polarization at three different incident angles. It is apparent that a sharp resonance is associated with each incident angle, and the resonance has a blue shift and slightly increasing bandwidth at a larger incident angle. We note that these resonances belong to the short-wavelength branch as shown in Fig. 3(a).

The SFG enhancement is most significant when both incident wavelengths match a certain QBIC resonance respectively. We first choose one incident plane wave with a fixed wavelength of $\lambda_1$=1490.665nm at the incident angle of 2°. The second incident beam has an incident angle of 3° while its wavelength $\lambda_2$ is scanned continuously. Both the incident plane waves are assumed to have an electric field amplitude of $1\times10^6$V/m, corresponding to an intensity of 0.133MW/cm$^2$ in vacuum. The SFG is calculated using the FEM method by only considering the $d_{33}$ value of $LiNbO_3$ as -41.7pm/V[35]. This simplification is valid because $d_{33}$ is one order of magnitude higher than other components and is the dominant factor in the second-order nonlinear process. As shown in Fig. 4(b), the SFG is most significant when $\lambda_2$ is tuned as 1478.818nm which is exactly the QBIC resonance for the incident angle of 3°, leading to a SFG wavelength of $\lambda_1\lambda_2/(\lambda_2+\lambda_2) = 742.359$nm. In the SFG calculations, we assume a grating length of 1cm is used in the y direction to have a valid 2D grating structure, and the power at SFG frequency is calculated by an integral of its Poynting vector only at the lower output port. From the results one can see that the SFG efficiency when both input wavelengths match the QBIC wavelength is enhanced by a factor of $10^8$, compared to the SFG effect through a bare

LiNbO$_3$ thin film of the same thickness. If one aspires to have the enhanced SFG at a different target wavelength, one can keep the wavelength of λ$_1$ at 2° and simply tune the incident angle of λ$_2$, which will tune the QBIC resonance to a different wavelength value. For example, when the incident angle of λ$_2$ is increased to 5°, the QBIC resonance will switch to 1454.755nm and the enhanced SFG will be around 736.246nm now. The SFG results for this case are presented in Fig. 4(c).

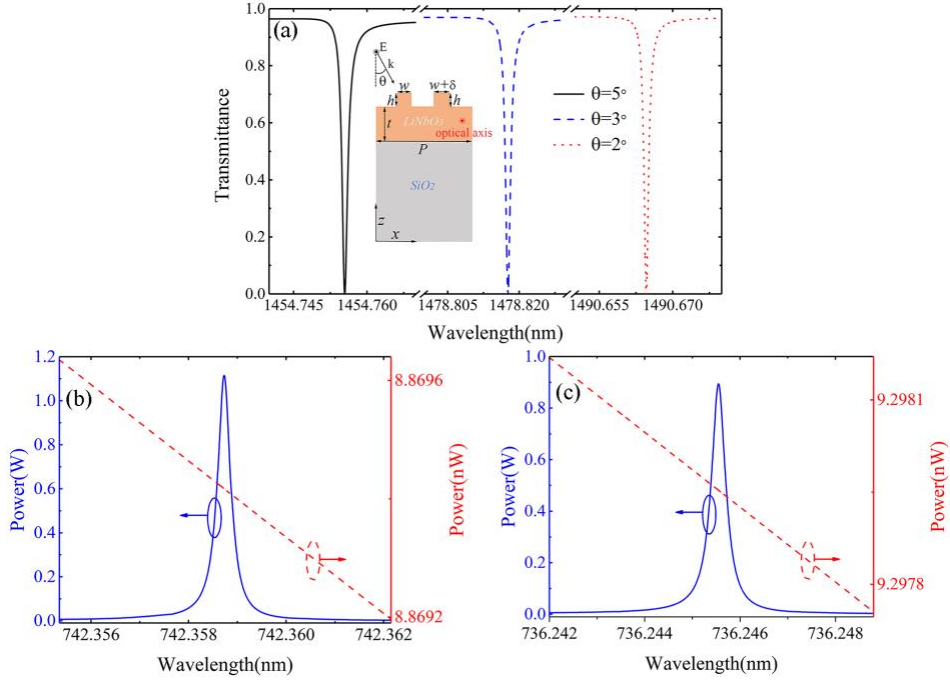

Fig.4. (a) Transmission spectra through the LiNbO$_3$ thin film binary grating structure at three different incident angles of 2°, 3° and 5°. The geometrical parameters are as follows: $P$=835nm, $t$=350nm, $h$=80nm, $w$=80nm, δ =5nm. (b) and (c) SFG spectra when one input wavelength is 1490.665nm at a fixed incident angle of 2° while the other input beam is fixed at 3° (b) and 5°(c) respectively while its wavelength is tuned. The red dashed lines correspond to the SFG through the bare LiNbO$_3$ thin film of the same thickness.

One of the requirements to achieve a large efficiency for nonlinear applications with multiple inputs is to have a large modal overlap between the input beams within the nonlinear medium to facilitate the interaction. For the SFG example, we are using the 1D QBIC resonances on the high-frequency branch in Fig. 3(d) for demonstration. As shown by the resonance mode distributions in Figs. 3(a) ~ (c), the modes exhibit strong similarities at different incident angles, which indicate that a large modal overlap can be achieved. The significantly enhanced SFG efficiency shown above supports this point. Since resonances on this branch exihibit larger Q factors at smaller incident angles, it is advantageous to use lower incident angles.

We note that all practical laser sources have certain bandwidths which may be larger than that of the QBIC, and it is only the frequency component of the QBIC resonance whose local electric field will be enhanced. So a laser source working in the continuous-wave mode with the central wavelength matching the QBIC resonance is preferred as the input for nonlinear applications. It may be challenging to achieve the spectral matching. Fortunately, this problem can be circumvented by simply tuning the incident angle, making use of the superior property of continuous distribution with the 1D QBIC effect.

## 4. Discussions and Conclusion

The most significant feature of the 1D QBIC resonances with the binary grating is that the QBICs can be supported continuously following the relation between the resonance wavelength and the incident angle governed by equation (1). As a result, one can achieve the QBIC resonance over a large spectrum by tuning the incident angle, and control the overall Q factor of these 1D QBIC by manipulating the degree of structural asymmetry. The continuous distribution of the QBIC resonances over a broad spectral range is a significant advantage over traditional QBIC resonances which can only occur around very few discrete positions. As a result, one can manipulate the light-matter interactions at any wavelength within the range, by simply choosing the proper incident angle. As an example, we have demonstrated in section 3 the enhancement of the SFG process with some spectral tunability by simply changing the incident angle of one input beam.

In summary, we have demonstrated in this work that a binary grating structure composed of two ridge arrays with the same period and slightly different ridge width located on a waveguide slab can be employed to support the 1D QBIC resonances along a continuous curve over a large spectral range in the ω~k space. The occurrence of the BIC/QBIC resonances at any wavelength over a broad spectral range for a structure with fixed geometry makes it possible to achieve the enhanced light-matter interactions with more freedom compared to the traditional BICs. We believe that these 1D QBICs can greatly promote many applications requiring multiple input wavelengths, and has great applications in general nonlinear optics. Furthermore, although we use the simple 1D grating structures to demonstrate the formation of the 1D QBIC resonances, the same methodology can be easily extended to more sophisticated 2D composite periodic elements or metasurface structures and to other spectral range of the electromagnetic spectrum to have enhanced interactions.